\documentclass[aps,floats]{revtex4}
\usepackage{amsmath,amssymb}
\usepackage{graphicx,epsfig}
\usepackage[greek,english]{babel}

\begin{document}
\bibliographystyle {plain}

\def\oppropto{\mathop{\propto}} 
\def\opsimeq{\mathop{\simeq}}
\def\opoverderline{\mathop{\overline}}
\def\operarrow{\mathop{\longrightarrow}}
\def\opsim{\mathop{\sim}}

\def\fig#1#2{\includegraphics[height=#1]{#2}}
\def\figx#1#2{\includegraphics[width=#1]{#2}}


\title{ Dissipative random quantum spin chain with boundary-driving and bulk-dephasing : \\
magnetization and current statistics in the Non-Equilibrium-Steady-State  } 


\author{ C\'ecile Monthus }
 \affiliation{Institut de Physique Th\'{e}orique, 
Universit\'e Paris Saclay, CNRS, CEA,
91191 Gif-sur-Yvette, France}

\begin{abstract}
The Lindblad dynamics with dephasing in the bulk and magnetization-driving at the two boundaries is studied for the quantum spin chain with random fields $h_j$ and couplings $J_j$ (that can be either uniform or random). In the regime of strong disorder in the random fields, or in the regime of strong bulk-dephasing, the effective dynamics can be mapped onto a classical Simple Symmetric Exclusion Process with quenched disorder in the diffusion coefficient associated to each bond. The properties of the corresponding Non-Equilibrium-Steady-State in each disordered sample between the two reservoirs are studied in detail by extending the methods that have been previously developed for the Symmetric Exclusion Process without disorder. Explicit results are given for the magnetization profile, for the two-point correlations, for the mean current and for the current fluctuations, in terms of the random fields and couplings defining the disordered sample.

\end{abstract}

\maketitle

In disordered quantum systems, 
the phenomenon of Anderson Localization (see the book \cite{50years} and references therein) 
or its generalization with interactions called
Many-Body-Localization ( see the recent reviews \cite{revue_huse,revue_altman,revue_vasseur,revue_imbrie,review_mblergo,review_rare} and references therein) is due to the coherent character of the unitary dynamics.
When these systems are not isolated anymore but become 'open' \cite{open}, it is essential to understand
whether the dissipation processes that tend to destroy the quantum coherence are able to eliminate the localization 
phenomenon. 
This issue has been analyzed recently in the context of random quantum spin chains
following some Lindblad dynamics \cite{horvat,garrahan_mbl,prosen_mbl,altman_mbl,znidaric_mblsubdiff,znidaric_mbldeph,huveneers_mblstep,c_lindbladboundary},
where it is very important to distinguish the various types of dissipation : if the dissipation occur only at the boundaries, the coherent dynamics in the bulk is sufficient to maintain the localization properties, while if the dissipation occurs everywhere in the bulk via dephasing, the localization phenomenon will be destroyed and it is interesting to characterize the properties of this dissipative dynamics in the presence of disorder.

In the field of quantum spin chains without disorder, the Lindblad dynamics has been much studied
to characterize the non-equilibrium transport properties  
\cite{casati_step,znidaric_deph,znidaric_trans,znidaric_heisen,hubbard,horvat,clark_heat,clark_step,prosen_step,dragi_trans}
with many exact solutions 
\cite {prosen_third,znidaric_solvable1,znidaric_solvable2,znidaric_solvable3,dragi_ex,dragi_twist,prosen_mpsolu,dragi_mps}.
The Lindblad framework for quantum systems also allows to make the link with the field of non-equilibrium classical stochastic processes described by Master Equations (see the review \cite{derrida} and references therein) : for instance the relaxation properties can be obtained from the spectrum of the Lindblad operator \cite{cai_barthel,kollath,znidaric_relaxation,garrahan_metastability,cai}, 
the large deviation formalism has been used to access the full-counting statistics 
\cite{garrahan_s,ates_s,hickey_s,genway_s,znidaric_slargedev,znidaric_sanomalous,prosen_s,pigeon},
the additivity principle has been tested \cite{znidaric_sadditivity}
and quantum fluctuation relations have been derived \cite{chetrite}.

In the present paper, our goal is to analyze the Lindblad dynamics of the XX quantum spin chain with random fields 
and couplings that can be either uniform or random, in the presence of dephasing in the bulk and in the presence of
magnetization-driving at the two boundaries in order to generate a Non-Equilibrium-Steady-State carrying a current.
 In the absence of bulk-dephasing,
this model has been found to keep its localized nature with a step magnetization profile and an exponentially decaying current with the system size \cite{c_lindbladboundary}. In the presence of bulk-dephasing, we obtain here that these localization properties are lost, as expected. 
We use the degenerate second-order perturbative approach in the XX-couplings $J_j$ developed previously either for strong bulk dephasing \cite{cai_barthel,znidaric_relaxation} or for strong disorder in the random fields \cite{prosen_mbl}. The effective dynamics can be then mapped onto a classical Simple Symmetric Exclusion Process with quenched disorder in the local diffusion coefficients. The methods that have been developed previously to study this classical stochastic model without quenched disorder (see the review \cite{derrida} and references therein) can be then adapted to characterize the Non-Equilibrium-Steady-State in each disordered sample and to obtain explicit results for the magnetizations, the correlations, and the two first cumulants of the integrated current.

The paper is organized as follows. 
In section \ref{sec_lindblad}, we introduce the notations for the Lindblad dynamics with boundary-driving and bulk-dephasing. In section \ref{sec_strong}, we focus on the regime of strong-disorder in the random fields or
on the regime of strong dephasing where the effective dynamics corresponds to a classical exclusion process
with random diffusion coefficients on the links. The properties of the corresponding Non-Equilibrium-Steady-State
in each disordered sample
are studied in the remaining of the paper, with explicit results for the magnetization profile and the averaged current
(section \ref{sec_magne}), for the two-point correlations (section \ref{sec_corre}) and for the current fluctuations
(section \ref{sec_fluctuat}). Our conclusions are summarized in section
\ref{sec_conclusion}.

\section{  Lindblad dynamics with boundary-driving and bulk-dephasing }

\label{sec_lindblad}

In this section, we describe the model for the Lindblad dynamics of the random field XX-chain 
with boundary-driving and bulk-dephasing.
As mentioned in the Introduction, the only difference with the previous work \cite{c_lindbladboundary}
is the presence of bulk dephasing that will completely change the physics.
We use the same notations to facilitate the comparison, but the two papers can also be read independently :
in the present section, we give a self-contained presentation of the model and of the notations.

\subsection{ Lindblad dynamics for the density matrix $\rho(t)$  }

 We consider the Lindblad dynamics for the density matrix $\rho(t)$ of the quantum chain of $N$ spins
\begin{eqnarray}
\frac{\partial \rho(t) }{\partial t} = -i [H,\rho ] 
+  {\cal D}^{Bulk}[\rho(t) ]
+ {\cal D}^{Left}[\rho(t) ] 
+ {\cal D}^{Right}[\rho(t) ] 
\label{dynlindblad}
\end{eqnarray}
The Hamiltonian contains random fields $h_j$ and
 XX-couplings $J_j$ (that can be either uniform or random)
\begin{eqnarray}
H&& = \sum_{j=1}^N \left[ h_j \sigma_j^z + J_j ( \sigma_j^x \sigma_{j+1}^x+\sigma_j^y \sigma_{j+1}^y ) \right]
 = \sum_{j=1}^N \left[ h_j \sigma_j^z
 + 2 J_j ( \sigma_j^+ \sigma_{j+1}^-+\sigma_j^- \sigma_{j+1}^+ ) \right]
\label{XXh}
\end{eqnarray}
It is possible to add couplings $J^z_j \sigma_i^z \sigma_{j+1}^z$,
but these couplings turn out to disappear at leading order in the strong-dephasing approximation \cite{cai_barthel,znidaric_relaxation}
 or in the strong-disorder approximation
\cite{prosen_mbl} 
that we will consider (see more details in section \ref{sec_per}).

The Bulk-dephasing operator acting with some amplitudes $\gamma_j$ (that can be taken uniform)
\begin{eqnarray}
 {\cal D}^{Bulk}[\rho] = \sum_{j=1}^N \gamma_{j} \left( \sigma_{j}^z\rho \sigma_{j}^z  -  \rho \right)
\label{dissidephasing}
\end{eqnarray}
tends to destroy off-diagonal elements with respect to the $\sigma^z$ basis.

The Left-Magnetization-driving
\begin{eqnarray}
 {\cal D}^{Left}[\rho] 
&& = \Gamma \frac{1+\mu}{2}  \left( \sigma_1^+ \rho \sigma_1^-  - \frac{1}{2} \sigma_1^- \sigma_1^+ \rho- \frac{1}{2} \rho  \sigma_1^- \sigma_1^+\right)
\nonumber \\
&& 
+\Gamma\frac{1-\mu}{2}  \left( \sigma_1^- \rho \sigma_1^+  - \frac{1}{2} \sigma_1^+ \sigma_1^- \rho- \frac{1}{2} \rho  \sigma_1^+ \sigma_1^-\right)
\label{dissiboundaryl}
\end{eqnarray}
tends to impose the magnetization $(\mu)$ on the first spin $\sigma_1$.
The Right-Magnetization-driving
\begin{eqnarray}
 {\cal D}^{Right}[\rho] 
&& = 
 \Gamma'\frac{1+\mu'}{2} \left( \sigma_N^+ \rho \sigma_N^-  - \frac{1}{2} \sigma_N^- \sigma_N^+ \rho- \frac{1}{2} \rho  \sigma_N^- \sigma_N^+\right)
\nonumber \\
&& 
+\Gamma'\frac{1-\mu'}{2}  \left( \sigma_N^- \rho \sigma_N^+  - \frac{1}{2} \sigma_N^+ \sigma_N^- \rho- \frac{1}{2} \rho  \sigma_N^+ \sigma_N^-\right)
\label{dissiboundaryr}
\end{eqnarray}
tends to impose the magnetization $(\mu')$ on the last spin $\sigma_N$.

When $\mu \ne \mu'$, the dynamics will converge at large times towards some stationary current-carrying non-equilibrium-steady-state $\rho^{ness}$
satisfying
\begin{eqnarray}
0=\frac{\partial \rho^{ness} }{\partial t} = -i [H,\rho^{ness} ] 
+  {\cal D}^{Bulk}[\rho^{ness} ]
+ {\cal D}^{Left}[\rho^{ness} ] 
+ {\cal D}^{Right}[\rho^{ness} ] 
\label{defness}
\end{eqnarray}

In the absence of disorder, the Lindblad dynamics of the pure XX chain with boundary-driving and bulk-dephasing has been studied in detail in various regimes (see \cite{horvat,znidaric_trans,znidaric_sadditivity,znidaric_solvable1,znidaric_solvable3} and references therein).
In the present paper, we will thus focus only on the disordered model.

\subsection{ Ladder Lindbladian for the ket $\vert \rho (t) \rangle $  }

To have a clearer picture of the Lindblad dynamics, it will be useful to introduce its spectral decomposition 
into eigenvalues and the corresponding Left and Right eigenvectors (see section \ref{sec_spectral} ). 
But in order to be able to use the very convenient  bra-ket notations, 
one needs first to 'vectorize' the density matrix as we now recall.

The density matrix $\rho(t)$ of the chain of $N$ spins can be expanded in the $\sigma^z$ basis 
\begin{eqnarray}
\rho(t) = \sum_{S_1=\pm 1} ...
 \sum_{S_N=\pm 1}  \sum_{T_1=\pm 1} ... \sum_{T_N=\pm 1} 
  \rho_{S_1,..,S_N;T_1,...,T_N} (t) \vert S_1,...,S_N \rangle \langle T_1,...,T_N \vert
\label{rhoexpcoefs}
\end{eqnarray}
in terms of the $4^N$ coefficients
\begin{eqnarray}
 \rho_{S_1,..,S_N;T_1,...,T_N } (t) = \langle  S_1,...,S_N \vert \rho(t) \vert T_1,...,T_N \rangle
\label{rhocoefs}
\end{eqnarray}
It is technically convenient to 'vectorize' the density {\it matrix } 
of the {\it spin chain } \cite{znidaric_relaxation,znidaric_sadditivity,jakob,savona,cirac,c_lindbladboundary},
i.e. to consider that these $4^N$ coefficients
are the components of a { \it ket } describing the state of a { \it spin ladder }
\begin{eqnarray}
\vert \rho (t) \rangle = \sum_{S_1=\pm 1} ... \sum_{S_N=\pm 1}  \sum_{T_1=\pm 1} ... \sum_{T_N=\pm 1} 
  \rho_{S_1,..,S_N;T_1,...,T_N} (t) \vert S_1,...,S_N \rangle \otimes \vert T_1,...,T_N \rangle
\label{rholadder}
\end{eqnarray}

The Lindbladian governing the dynamics of the ket $\vert \rho (t) \rangle $
\begin{eqnarray}
\frac{\partial \vert \rho (t) \rangle }{\partial t} = {\cal L}
\vert \rho (t) \rangle
\label{dynlindbladladder}
\end{eqnarray}
reads in this ladder formulation
\begin{eqnarray}
 {\cal L}
&& = -i  \sum_{j=1}^N \left[ h_j \sigma_j^z 
 + 2 J_j ( \sigma_j^+ \sigma_{j+1}^-+\sigma_j^- \sigma_{j+1}^+ ) \right] 
+ i  \sum_{j=1}^N \left[ h_j \tau_j^z 
 + 2 J_j ( \tau_j^+ \tau_{j+1}^-+\tau_j^- \tau_{j+1}^+ ) \right] 
\nonumber \\
&& -  \sum_{j=1}^N \gamma_{j} \left( 1- \sigma_{j}^z \tau_{j}^z  \right)
 \nonumber \\
&&
+ \Gamma \left( \frac{1+\mu}{2}  \sigma_1^+ \tau_1^+ + \frac{1-\mu}{2}  \sigma_1^- \tau_1^-  \right)
-  \frac{ \Gamma}{2}
 + \frac{ \Gamma \mu}{4} (\sigma_1^z + \tau_1^z )
  \nonumber \\
&&
+ \Gamma' \left( \frac{1+\mu'}{2}  \sigma_N^+ \tau_N^+ + \frac{1-\mu'}{2}  \sigma_N^- \tau_N^-  \right)
-  \frac{ \Gamma' }{2}
 + \frac{ \Gamma' \mu'}{4} (\sigma_N^z + \tau_N^z )
\label{dissiboundaryrightladder}
\end{eqnarray}

\subsection { Spectral Decomposition into eigenvalues and eigenstates }

\label{sec_spectral}

In this section, we describe the spectral decomposition of the Lindbladian when it is diagonalisable
(when it is not diagonalisable, one has to use instead the decomposition into Jordan blocks,
but we will not need to consider this complication here).

It is convenient to use the bra-ket notations
to denote the Right and Left eigenvectors
associated to the $4^N$ eigenvalues $\lambda_n$
\begin{eqnarray}
 {\cal L}  \vert \lambda_n^R \rangle && = \lambda_n  \vert \lambda_n^R \rangle
\nonumber \\
\langle\lambda_n^L \vert {\cal L}  && = \lambda_n \langle\lambda_n^L \vert
\label{spectraleigen}
\end{eqnarray}
with the orthonormalization
\begin{eqnarray}
  \langle\lambda_n^L \vert  \lambda_m^R \rangle = \delta_{nm}
\label{orthonorm}
\end{eqnarray}
and the identity decomposition
\begin{eqnarray}
1 = \sum_{n=0}^{4^N-1}   \vert \lambda_n^R\rangle  \langle\lambda_n^L \vert 
\label{identity}
\end{eqnarray}
The spectral decomposition of the Lindbladian
\begin{eqnarray}
 {\cal L} =  \sum_{n=0}^{4^N-1}  \lambda_n   \vert \lambda_n^R \rangle  \langle\lambda_n^L \vert 
\label{spectral}
\end{eqnarray}
yields the solution for the dynamics in terms of the initial condition at $t=0$ 
\begin{eqnarray}
\vert \rho(t) \rangle=  \sum_{n=0}^{4^N-1}  e^{\lambda_n t}   \vert 
 \lambda_n^R \rangle  \langle\lambda_n^L 
 \vert \rho(t=0) \rangle
\label{spectralrlax}
\end{eqnarray}

The trace of the density matrix $\rho(t)$ corresponds  in the Ladder Formulation to
\begin{eqnarray}
Trace(\rho(t) ) =\sum_{S_1=\pm 1} ... \sum_{S_N=\pm 1} 
  \rho_{S_1,..,S_N;S_1,...,S_N} (t)
=  \sum_{S_1=\pm 1} ... \sum_{S_N=\pm 1} \langle  S_1,..,S_N \vert \otimes \langle S_1,...,S_N
\vert \rho (t) \rangle
\label{tracerholader}
\end{eqnarray}
The conservation of $ Trace(\rho(t) )$ by the dynamics  means that the vanishing eigenvalue
\begin{eqnarray}
\lambda_0=0
\label{zeroeigen}
\end{eqnarray}
is always present in the spectrum and associated to the Left eigenvector
\begin{eqnarray}
  \langle\lambda_0^L\vert  = \sum_{S_1=\pm 1} ... \sum_{S_N=\pm 1}
 \langle  S_1,..,S_N \vert \otimes \langle S_1,...,S_N
\vert
\label{lefteigenzero}
\end{eqnarray}
For the present model where the steady-state is non-degenerate (see \cite{prosen_uniqueness} and references therein
for the general strategy to prove the uniqueness of the steady state),
the steady state $\rho^{ness} $ of Eq. \ref{defness}  towards which any initial condition will converges via Eq. \ref{spectralrlax}
corresponds to the unique Right Eigenvector associated to the vanishing eigenvalue $\lambda_0=0 $
\begin{eqnarray}
\vert \rho(t\to +\infty ) \rangle=    \vert \lambda_0^R \rangle = \vert \rho^{ness} \rangle 
\label{righteigenzero}
\end{eqnarray}
The other $(4^N-1)$ eigenvalues $\lambda_{n \ne 0}$ with negative real parts
describe the relaxation towards this steady state.

\section{ Effective Lindbladian for strong disorder or strong dephasing }

\label{sec_strong}

\subsection{ Perturbation in the couplings $J_j $  }

\label{sec_per}

In this section, we consider that the terms of the Lindbladian containing the couplings $J_j$
\begin{eqnarray}
 {\cal L}^{per}  
&& =i  \sum_{j=1}^{N-1}  2 J_j ( \tau_j^+ \tau_{j+1}^-+\tau_j^- \tau_{j+1}^+ 
- \sigma_j^+ \sigma_{j+1}^--\sigma_j^- \sigma_{j+1}^+)  
\label{per}
\end{eqnarray}
can be treated perturbatively with respect to the other terms of the Lindbladian that do not couple the rungs of the ladder
\begin{eqnarray}
 {\cal L}^{unper}  
&& = \sum_{j=1}^N  {\cal L}^{unper}_j  
\label{unper}
\end{eqnarray}
The Lindbladians associated to the rungs of the bulk $j=2,..,N-1$ read
\begin{eqnarray}
   {\cal L}^{unper}_j  =  i   h_j (\tau_j^z -\sigma_{j}^z  )-   \gamma_{j} \left( 1- \sigma_{j}^z \tau_{j}^z  \right) 
\label{unperj}
\end{eqnarray}
while for the two end-spins, they contain the additional contribution of the boundary-driving
\begin{eqnarray}
   {\cal L}^{unper}_{j=1}  = i   h_1 (\tau_1^z -\sigma_1^z  ) -   \gamma_1 \left( 1- \sigma_1^z \tau_1^z  \right) 
+ \Gamma \left( \frac{1+\mu}{2}  \sigma_1^+ \tau_1^+ + \frac{1-\mu}{2}  \sigma_1^- \tau_1^-  \right)
-  \frac{ \Gamma}{2}
 + \frac{ \Gamma \mu}{4} (\sigma_1^z + \tau_1^z )
\label{unper1}
\end{eqnarray}
and
\begin{eqnarray}
   {\cal L}^{unper}_{j=N}  =  i   h_N (\tau_N^z -\sigma_N^z  )-   \gamma_N \left( 1- \sigma_N^z \tau_N^z  \right) 
+ \Gamma' \left( \frac{1+\mu'}{2}  \sigma_N^+ \tau_N^+ + \frac{1-\mu'}{2}  \sigma_N^- \tau_N^-  \right)
-  \frac{ \Gamma' }{2}
 + \frac{ \Gamma' \mu'}{4} (\sigma_N^z + \tau_N^z )
\label{unpern}
\end{eqnarray}

In the absence of boundary-drivings, 
this type of perturbation theory has been developed previously for the pure $XXZ$ chain without fields
as a 'strong dissipation' approximation \cite{cai_barthel,znidaric_relaxation},
and in XXZ-chain with random fields as a 'strong disorder' approximation \cite{prosen_mbl},
in order to analyze the relaxation properties towards the trivial maximally mixed steady-state.
Note that in both cases, the $J^z$-coupling turns out to disappear
at leading order in this perturbation theory \cite{cai_barthel,prosen_mbl},
and this is why we have chosen to consider here the case $J^z_i=0$ from the very beginning (Eq \ref{XXh})
in order to simplify the presentation.
In the following, we describe how the perturbation theory developed in \cite{cai_barthel,znidaric_relaxation,prosen_mbl}
has to be adapted to the presence of the boundary-drivings.

\subsection { Spectral decomposition of ${\cal L}^{unper}_{1} $  }

The Lindbladian ${\cal L}^{unper}_{1} $ of Eq. \ref{unper1} can be rewritten in terms of its spectral decomposition (Eq \ref{spectral})
\begin{eqnarray}
 {\cal L}^{unper}_{1}  =  \sum_{n=0}^{3}  \lambda_{1,n}   \vert \lambda_{1,n}^R \rangle  \langle\lambda_{1,n}^L \vert 
\label{spectrall1unper}
\end{eqnarray}
where the four eigenvalues $\lambda_{1,n}  $ 
and the corresponding Left and Right Eigenvectors written in the basis $(\sigma_1^z,\tau^z_1)$ are :

(0) The eigenvalue $\lambda_{1,n=0} =0$ is associated to 
\begin{eqnarray}
  \langle  \lambda_{1,n=0}^{L} \vert  && =   \langle ++ \vert + \langle -- \vert
\nonumber \\
    \vert \lambda_{1,n=0}^{R}  \rangle && = \frac{1+\mu}{2} \vert ++\rangle +\frac{1-\mu}{2} \vert -- \rangle  
\label{lindbladspin1n0}
\end{eqnarray}

(1) The eigenvalue $\lambda_{1,n=1}=-\Gamma$ is associated to 
\begin{eqnarray}
  \langle \lambda_{1,n=1}^{L} \vert  && =  \frac{1-\mu}{2} \langle ++ \vert  - \frac{1+\mu}{2} \langle--\vert 
\nonumber \\
    \vert \lambda_{1,n=1}^{R}  \rangle && =  \vert ++\rangle  - \vert --\rangle
\label{lindbladspin1n1}
\end{eqnarray}

(2) The eigenvalue $\lambda_{1,n=2}=-\frac{\Gamma}{2}-2 \gamma_1+i 2h_1$ is associated to 
\begin{eqnarray}
  \langle\lambda_{1,n=2}^{L}  \vert  && = \langle -+ \vert  
\nonumber \\
    \vert \lambda_{1,n=2}^{R} \rangle && = \vert -+\rangle  
\label{lindbladspin1n2}
\end{eqnarray}

(4) The eigenvalue $\lambda_{1,n=3}=-\frac{\Gamma}{2}-2 \gamma_1-i 2h_1$ is associated to 
\begin{eqnarray}
  \langle\lambda_{1,n=3}^{L} \vert  && = \langle +- \vert  
\nonumber \\
    \vert \lambda_{1,n=3}^{R}  \rangle && = \vert +-\rangle  
\label{lindbladspin1n3}
\end{eqnarray}

\subsection { Spectral decomposition of ${\cal L}^{unper}_N $   }

Similarly, the Lindbladian ${\cal L}^{unper}_{N} $ of Eq. \ref{unpern} can be rewritten in terms of its spectral decomposition (Eq \ref{spectral})
\begin{eqnarray}
 {\cal L}^{unper}_{N}  =  \sum_{m=0}^{3}  \lambda_{N,m}   \vert \lambda_{N,m}^R \rangle  \langle\lambda_{N,m}^L \vert 
\label{spectrallnunper}
\end{eqnarray}
where the four eigenvalues $\lambda_{1,m}  $ 
and the corresponding Left and Right Eigenvectors written in the basis $(\sigma_N^z,\tau^z_N)$ are:

(0) The eigenvalue $\lambda_{N,m=0} =0$ is associated to 
\begin{eqnarray}
  \langle \lambda_{N,m=0}^L  \vert  && =   \langle ++ \vert + \langle -- \vert
\nonumber \\
    \vert  \lambda_{N,m=0}^R \rangle && = \frac{1+\mu'}{2} \vert ++\rangle +\frac{1-\mu'}{2} \vert -- \rangle  
\label{lindbladspin2n0}
\end{eqnarray}

(1) The eigenvalue $\lambda_{N,m=1}=-\Gamma'$ is associated to 
\begin{eqnarray}
  \langle  \lambda_{N,m=1}^L\vert  && =  \frac{1-\mu'}{2} \langle ++ \vert  - \frac{1+\mu'}{2} \langle--\vert 
\nonumber \\
    \vert  \lambda_{N,m=1}^R \rangle && =  \vert ++\rangle  - \vert --\rangle
\label{lindbladspin2n1}
\end{eqnarray}

(2) The eigenvalue $\lambda_{N,m=2}=-\frac{\Gamma'}{2}-2 \gamma_N+i 2h_N$ is associated to 
\begin{eqnarray}
  \langle\lambda_{N,m=2}^L\vert  && = \langle -+ \vert  
\nonumber \\
    \vert  \lambda_{N,m=2}^R\rangle && = \vert -+\rangle  
\label{lindbladspin2n2}
\end{eqnarray}

(4) The eigenvalue $\lambda_{N,m=3}=-\frac{\Gamma'}{2}-2 \gamma_N-i 2h_N$ is associated to 
\begin{eqnarray}
  \langle \lambda_{N,m=3}^L\vert  && = \langle +- \vert  
\nonumber \\
    \vert  \lambda_{N,m=3}^R\rangle && = \vert +-\rangle  
\label{lindbladspin2n3}
\end{eqnarray}

\subsection { Spectral decomposition of ${\cal L}^{unper}_j $ for $j=2,..,N-1$    }

The Lindbladian operator ${\cal L}^{unper}_j  $ (Eq. \ref{unperj}) in the bulk $j=2,..,N-1$
is diagonal in the $(\sigma_j^z,\tau_j^z)$ basis : it is thus more convenient to write
its spectral decomposition (Eq. \ref{spectral}) as
\begin{eqnarray}
 {\cal L}^{unper}_{j}  =  \sum_{S_j=\pm,T_j=\pm} \lambda_{j ,S_j,T_j}    \vert S_j,T_j  \rangle  \langle  S_j,T_j \vert 
\label{spectralljunper}
\end{eqnarray}
 with the eigenvalues
\begin{eqnarray}
   \lambda_{j ,S_j,T_j} =  i   h_j (T_j - S_j )-   \gamma_{j} \left( 1- S_j,T_j  \right) 
\label{ljsjtj}
\end{eqnarray}
So here the vanishing eigenvalue is degenerate twice
\begin{eqnarray}
   \lambda_{j ,+,+} =\lambda_{j ,-,-} =0
\label{lj0}
\end{eqnarray}
while the two others eigenvalue read
\begin{eqnarray}
    \lambda_{j ,+,-}    =-2 \gamma_j  - i  2 h_j 
\nonumber \\
  \lambda_{j ,-,+}    =-2 \gamma_j  + i  2 h_j 
\label{ljnon0}
\end{eqnarray}

\subsection { Spectral decomposition of ${\cal L}^{unper}$     }

The unperturbed Lindbladian of Eq \ref{unper} is the sum of the independent Lindbladians discussed above.
So its eigenvalues are simply given by the sum of eigenvalues 
\begin{eqnarray}
 \lambda^{unper}_{(1,n);(j,S_j,T_j);(N,m)} && =\lambda_{1,n}  + \sum_{j=2}^{N-1} \lambda_{j ,S_j,T_j}+ \lambda_{N,m}
\label{unpertwonm}
\end{eqnarray}
while the left and right eigenvectors are given by the corresponding tensor-products.

In particular the vanishing eigenvalue $\lambda^{unper}=0$ is very degenerate,
and the corresponding subspace of dimension $2^{N-2}$ is described by the projector
\begin{eqnarray}
{\cal P}_0 && = \sum_{S_2,...,S_{N-1}}  
\vert \lambda_{1,n=0}^{R} \rangle \otimes_{j=2}^{N-1} \vert S_j,T_j=S_j \rangle \otimes \vert  \lambda_{N,m=0}^R \rangle
\langle\lambda_{1,n=0}^{L} \vert \otimes_{j=2}^{N-1} \langle S_j,T_j =S_j\vert \otimes \langle  \lambda_{N,m=0}^L \vert
\label{proj0}
\end{eqnarray}

\subsection{ Perturbation theory within the degenerate subspace associated to $\lambda^{unper}=0$ }

Within the degenerate subspace of dimension $2^{N-2}$ associated to $\lambda^{unper}=0$,
the effective dynamics is described by the operator obtained by the second-order perturbation formula \cite{cai_barthel,znidaric_relaxation,prosen_mbl}
\begin{eqnarray}
W \equiv {\cal L}^{(2^dorder)} _{\lambda^{unper}=0}=  {\cal P}_0 {\cal L}^{per} (1-{\cal P}_0 )\frac{1}{0- {\cal L}^{unper}  }
(1-{\cal P}_0 ){\cal L}^{per} {\cal P}_0
\label{Leff2emeordre}
\end{eqnarray}

The action of the perturbation $ {\cal L}^{per}   $ of Eq. \ref{per}
on the Left-Eigenvectors 
\begin{eqnarray}
&& \langle\lambda_{1,n=0}^{L} \vert \otimes_{j=2}^{N-1} \langle S_j=\eta_j,T_j =\eta_j \vert \otimes \langle  \lambda_{N,m=0}^L \vert
 {\cal L}^{per}  
\nonumber \\
&&  = i 2 J_1 \left( \delta_{\eta_2=+} -   \delta_{\eta_2=-} e^{-s}  \right) 
\left(  \langle\lambda_{1,n=3}^{L} \vert \otimes \langle S_2=-,T_2 =+\vert
 - \langle \lambda_{1,n=2}^{L} \vert \otimes \langle S_2=+,T_2 =-\vert\right) 
\nonumber \\
&& \otimes_{j=3}^{N-1} \langle S_j=\eta_j,T_j =\eta_j\vert \otimes \langle  \lambda_{N,m=0}^R \vert
\nonumber \\
&& 
+ i  \sum_{k=2}^{N-2}  2 J_k \delta_{\eta_{k+1}=-\eta_k}
\langle\lambda_{1,n=0}^{L} \vert \otimes_{j=2}^{k-1} \langle S_j=\eta_j,T_j =\eta_j \vert 
\nonumber \\
&& \left( \langle S_k=\eta_k,T_k=-\eta_k \vert \otimes \langle S_{k+1}=-\eta_k,T_{k+1}=\eta_k \vert
  -   \langle S_k=-\eta_k,T_k=\eta_k \vert \otimes \langle S_{k+1}=\eta_k,T_{k+1}=-\eta_k \vert\right)
\nonumber \\
&& \otimes_{j=k+2}^{N-1} \langle S_j=\eta_j,T_j =\eta_j\vert
\otimes \langle  \lambda_{N,m=0}^L \vert
\nonumber \\
&& + i  2 J_{N-1}  
\langle\lambda_{1,n=0}^{R} \vert \otimes_{j=2}^{N-2} \langle S_j=\eta_j,T_j =\eta_j\vert \otimes 
\nonumber \\
&&\left( \delta_{\eta_{N-1}=+} -   \delta_{\eta_{N-1}=-}  \right) 
  \left( \langle S_{N-1}=+,T_{N-1} =-\vert \otimes \langle  \lambda_{N,m=2}^L \vert 
- \langle S_{N-1}=-,T_{N-1} =+\vert \otimes \langle  \lambda_{N,m=3}^L \vert  \right)
\label{proj0l}
\end{eqnarray}
and on the right eigenvectors
\begin{eqnarray}
&&  {\cal L}^{per}   \vert \lambda_{1,n=0}^{R} \rangle \otimes_{j=2}^{N-1} \vert S_j=\eta_j',T_j=\eta_j' \rangle \otimes \vert  \lambda_{N,m=0}^R \rangle
\nonumber \\
&& = i   2 J_1 \left( \delta_{\eta_2'=+} \frac{1-\mu}{2}  -   \delta_{\eta_2'=-} e^{s} \frac{1+\mu}{2}  \right) 
\left(  \vert \lambda_{1,n=3}^{R} \rangle \otimes \vert S_2=-,T_2 =+\rangle - \vert \lambda_{1,n=2}^{R} \rangle \otimes \vert S_2=+,T_2 =- \rangle\right) 
\nonumber \\
&&\otimes_{j=3}^{N-1} \vert S_j=\eta_j',T_j=\eta_j' \rangle \otimes \vert  \lambda_{N,m=0}^R \rangle
\nonumber \\
&& + i  \sum_{k=1}^{N-2}  2 J_k\delta_{\eta_{k+1}'=-\eta_k'}
\vert \lambda_{1,n=0}^{R} \rangle \otimes_{j=2}^{k-1} \vert S_j=\eta_j',T_j=\eta_j' \rangle 
\nonumber \\
&& \left( \vert S_k=\eta_k',T_k=-\eta_k' \rangle \otimes \vert S_{k+1}=-\eta_k',T_{k+1}=\eta_k' \rangle 
-\vert S_k=-\eta_k',T_k=\eta_k' \rangle \otimes \vert S_{k+1}=\eta_k',T_{k+1}=-\eta_k'  \rangle \right)
\nonumber \\
&& \otimes_{j=k+2}^{N-1} \vert S_j=\eta_j',T_j=\eta_j' \rangle 
\otimes \vert  \lambda_{N,m=0}^R \rangle
\nonumber \\
&& + i   2 J_{N-1}\vert \lambda_{1,n=0}^{R} \rangle \otimes_{j=2}^{N-2} \vert S_j=\eta_j',T_j=\eta_j' \rangle
\left( \delta_{\eta_{N-1}'=+} \frac{1-\mu'}{2} -   \delta_{\eta_{N-1}'=-} \frac{1+\mu'}{2}  \right)  
\nonumber \\
&&
 \left( \vert S_{N-1}=+,T_{N-1} =- \rangle \otimes \vert  \lambda_{N,m=2}^R \rangle 
- \vert S_{N-1}=-,T_{N-1} =+ \rangle \otimes \vert \lambda_{N,m=3}^R \rangle \right)
\label{proj0r}
\end{eqnarray}
determine the intermediate unperturbed states that appear in the perturbative formula of Eq. \ref{Leff2emeordre}.
Using the corresponding unperturbed eigenvalues of Eq. \ref{unpertwonm} that appear in the denominators,
one finally obtains that the effective operator $W$ (Eq. \ref{Leff2emeordre}) acting on the $(N-2)$ spins $(S_2,..,S_{N-1})$ labeling the
degenerate subspace of Eq. \ref{proj0}
reads in terms of Pauli matrices
\begin{eqnarray}
W && = D_{1,2} \left( 
   \frac{1+\mu}{2} \sigma_2^+ 
 +  \frac{1-\mu}{2} \sigma_2^-
- \frac{1- \mu \sigma_2^z}{2}
\right)
\nonumber \\
&&
+  \sum_{k=2}^{N-2} D_{k,k+1} \left(
\sigma_k^{+}\sigma_{k+1}^{-} +\sigma_k^{-}\sigma_{k+1}^{+} 
- \frac{1- \sigma_k^z \sigma_{k+1}^z}{2} )   \right)
\nonumber \\
&&
+ D_{N-1,N}
\left(   \frac{1+\mu'}{2}  \sigma_{N-1}^+
 +   \frac{1-\mu'}{2} \sigma_{N-1}^-
-  \frac{1-\mu'\sigma_{N-1}^z}{2} 
  \right) 
\label{wmaster}
\end{eqnarray}
where we have introduced the notations
\begin{eqnarray}
D_{k,k+1} && \equiv  \frac{ 4 J_k^2(\gamma_k+\gamma_{k+1}) }{ (\gamma_k+\gamma_{k+1})^2+(h_k-h_{k+1})^2}  \ \ { \rm for} \ \ k=2,..,N-2
\nonumber \\
D_{1,2} && \equiv \frac{ 4 J_1^2 ( \Gamma+4 (\gamma_1+\gamma_2)) }
{ \left(\frac{\Gamma}{2}+2 (\gamma_1+\gamma_2)   \right)^2+4 (h_1-h_2)^2}
 \nonumber \\
D_{N-1,N} &&  \equiv \frac{ 4 J_{N-1}^2 ( \Gamma'+4 (\gamma_{N-1}+\gamma_N))}
{ \left(\frac{\Gamma'}{2}+2 (\gamma_{N-1}+\gamma_N)   \right)^2+4 (h_{N-1}-h_N)^2}  
\label{notaD}
\end{eqnarray}

\subsection{ Validity of this perturbative approach }

The above approach is consistent if the bulk diffusion coefficients $D_{k,k+1} $ obtained by this
 second-order perturbation theory are indeed small with respect to the dephasing 
coefficients $\gamma_j$ appearing in the real parts of the unperturbed eigenvalues of Eq. \ref{ljnon0}.

For the pure model with homogeneous couplings $J_k=J$ and without random fields $h_k=0$, 
 where the bulk diffusion coefficient of Eq. \ref{notaD} becomes \cite{cai_barthel,znidaric_relaxation}
\begin{eqnarray}
D^{pure}=\frac{2 J^2}{\gamma} 
\label{Dpure}
\end{eqnarray}
the approximation is thus valid for strong bulk dephasing $\gamma \gg \vert J \vert$ \cite{cai_barthel,znidaric_relaxation}.

For our present disordered model,  the bulk diffusion coefficients of Eq. \ref{notaD}
become at leading order in the limit of strong disorder in the random fields $h_k$ \cite{prosen_mbl}
\begin{eqnarray}
D^{StrongDisorder}_{k,k+1} \simeq  \frac{ 4 J_k^2(\gamma_k+\gamma_{k+1}) }{ (h_k-h_{k+1})^2} 
\label{DStrongDisorder}
\end{eqnarray}
so here the approximation remains valid for $(h_k-h_{k+1} )^2 \gg J_k^2$ \cite{prosen_mbl}
for arbitrary dephasing coefficients $\gamma_k$, as long as they do not vanish.
Indeed when the bulk-dephasing is absent $\gamma_k=0$, the physics is of course completely different
as recalled in the Introduction
and another strong disorder approach is appropriate  \cite{c_lindbladboundary}.

From this discussion, it is clear that the perturbative approach which has been described
either as a strong dephasing approximation  \cite{cai_barthel,znidaric_relaxation}
or as a strong disorder approximation \cite{prosen_mbl} can be equivalently summarized
as a weak-coupling approximation 
\begin{eqnarray}
 J_k^2 \ll  (\gamma_k+\gamma_{k+1})^2+(h_k-h_{k+1})^2 
\label{Dweakcoupling}
\end{eqnarray}
where the couplings $J_k$ have to be weak with respect to the global effect of dephasing and random-field disorder.

To shed further light on the physical meaning of this approximation, 
it is also useful to interpret the perturbative approach as the decomposition of the Lindblad dynamics 
into two regimes  \cite{cai_barthel,znidaric_relaxation,prosen_mbl} :

(i) at short times $t \leq {\rm max} (\frac{1}{2 \gamma_j})$, the main effect of the Lindblad dynamics 
is to suppress the off-diagonal components as a consequence of dephasing on each site $j$ of the bulk :
 the convergence towards the diagonal elements associated to the degenerate zero-eigenvalue (Eq \ref{lj0})
is described by the non-zero eigenvalues $(-2 \gamma_j  \pm i  2 h_j )$ of Eq. \ref{ljnon0}.

(ii) then for larger times $ t \geq ( {\rm max} \frac{1}{2 \gamma_j})$, 
the effective dynamics between the remaining diagonal components of the
density matrix is described by the operator $W$ obtained above that takes into account the couplings between the sites.

\subsection{ Summary : mapping onto a classical exclusion process with disorder }

Let us now summarize the output of the above calculations.
The ket $\vert \rho (t)\rangle$ of the spin ladder of length $N$ with an Hilbert space of dimension $4^N$
has been projected onto the ket $ \vert P(t) \rangle$ of a spin chain of $(N-2)$ spins with an Hilbert space 
of dimension $2^{N-2}$ that represents the 
the diagonal elements
\begin{eqnarray}
\langle S_2,..,S_{N-1} \vert   P(t) \rangle = 
\langle\lambda_{1,n=0}^{L} \vert \otimes_{j=2}^{N-1} \langle S_j,T_j =S_j\vert \otimes \langle  \lambda_{N,m=0}^L \vert  \rho(t) \rangle
\label{ketp}
\end{eqnarray}

The Lindbladian that was acting on the ket $\vert \rho (t)\rangle$ 
has been projected onto the effective operator $W$ of Eq. \ref{wmaster}
that governs the dynamics of the ket $ \vert P(t) \rangle$
\begin{eqnarray}
\frac{\partial \vert P_t \rangle }{\partial t} = W \vert P_t \rangle
\label{master}
\end{eqnarray}
The spectral decomposition
\begin{eqnarray}
 W = \sum_{n=0}^{2^{N-2}-1} w_n \vert w_n^R \rangle\langle w_n^L \vert
\label{wspec}
\end{eqnarray}
that allows to rewrite the solution of the dynamics as
\begin{eqnarray}
 \vert P_t \rangle =  \sum_{n=0}^{2^{N-2}-1} e^{w_n } \vert w_n^R \rangle\langle w_n^L \vert P_{t=0} \rangle 
\label{psol}
\end{eqnarray}
has the same properties as the spectral decomposition of the Lindbladian :
the vanishing eigenvalue $w_{n=0}=0$ is associated to the Left Eigenvector
\begin{eqnarray}
 \langle w_{n=0}^L \vert = \sum_{S_2,..,S_{N-1} } \langle S_2,..,S_{N-1} \vert
\label{w0left}
\end{eqnarray}
that encodes the conservation of probability
\begin{eqnarray}
  \sum_{S_2,..,S_{N-1} } \langle S_2,..,S_{N-1} \vert P_t \rangle =1
\label{probamaster}
\end{eqnarray}
while the corresponding Right Eigenvector $\vert w_n^R \rangle $ corresponds to the non-equilibrium steady state
towards which any initial condition converges
\begin{eqnarray}
 \vert P_{t \to +\infty} \rangle =   \vert w_{n=0}^R \rangle
\label{w0right}
\end{eqnarray}
The other modes $n \ne 0$ describe the relaxation towards this steady state.

In \cite{cai_barthel,znidaric_relaxation,prosen_mbl}, the operator $W$ of Eq. \ref{wmaster} was written as minus the quantum Heisenberg ferromagnetic Hamiltonian
\begin{eqnarray}
- W && = H^{eff} =  \sum_{k=1}^{N-1} D_{k,k+1} \left(
 \frac{ 1- {\vec \sigma_k }.{\vec \sigma_{k+1} } }{2}    \right)
\label{wmasterferro}
\end{eqnarray}
to derive various consequences. In our present case, we will keep 
the writing of Eq. \ref{wmaster} and interpret it as a 
classical Master Equation describing a 
Simple Symmetric Exclusion Process 
with quenched disorder in the local diffusion coefficients $D_{k,k+1}$ (Eq \ref{notaD}).
The pure Simple Symmetric Exclusion Process with uniform $D_{k,k+1}=1$
is one of the standard model in the field of non-equilibrium classical systems 
(see the review \cite{derrida} and references therein). 
The effects of quenched disorder on { \it totally or partially asymmetric } exclusion models
have been analyzed in \cite{harris,enaud,santen,igloi,juhasz}.
In our present case, it is very important to stress that the disorder is in the 
local diffusion coefficients $D_{k,k+1}$, but that the { \it symmetry between the jumps 
from $k$ to $(k+1)$ or from $(k+1)$ to $k$ } is maintained.
On the contrary, for the model with random hopping rates that do not satisfy this symmetry, 
there exists a random local force that build a random potential landscape with large barriers 
that will govern the transport properties \cite{santen,igloi,juhasz}, so that the physics is completely different.

\subsection{ Dynamics of observables }

In the remaining of the paper, we wish to study various observables within the effective dynamics
described by the operator $W$.
The average at time $t$ of the observable associated to the operator $A$
\begin{eqnarray}
\langle A\rangle_t = \sum_{S_2,..,S_{N-1} } \langle S_2,..,S_{N-1} \vert A \vert P_t \rangle = \langle w_{n=0}^L \vert A \vert P_t \rangle
\label{defaveraget}
\end{eqnarray}
evolves in time according to the dynamical equation
\begin{eqnarray}
\frac{\partial \langle A\rangle_t }{\partial t} =\langle w_{n=0}^L \vert A W \vert P_t \rangle = \langle w_{n=0}^L \vert [A, W] \vert P_t \rangle = \langle [A, W] \rangle_t
\label{dynA}
\end{eqnarray}
where the commutator has been introduced using the property  $\langle w_{n=0}^L \vert W =0 $ 
of the Left eigenvector.
In the following sections, we analyze the properties of the non-equilibrium-steady-state (NESS) in each disordered sample,
via the magnetizations, the correlations and the statistics of the current.

\section{ Local magnetizations and local currents }

\label{sec_magne}

\subsection{ Dynamics of the local magnetizations  }

The dynamics of the magnetization on site $j$ 
is described by Eq. \ref{dynA} for $A=\sigma_j^z$
\begin{eqnarray}
\frac{\partial \langle\sigma_j^z\rangle_t }{\partial t} = \langle [\sigma_j^z, W] \rangle_t = \langle {\cal I}_{j-1,j}  -  {\cal I}_{j,j+1}  \rangle_t
\label{dynm}
\end{eqnarray}
that involves the current operators associated to the bonds $(j,j+1)$
\begin{eqnarray}
{\cal I}_{j,j+1} = - [\sigma_j^z, D_{j,j+1} \left(
\sigma_j^{+}\sigma_{j+1}^{-} +\sigma_j^{-}\sigma_{j+1}^{+}  \right)] 
=  2 D_{j,j+1} \left(\sigma_j^{-}\sigma_{j+1}^{+} 
- \sigma_j^{+}\sigma_{j+1}^{-}  \right)
\label{currentop}
\end{eqnarray}

From the definition of the average in Eq. \ref{defaveraget}, one obtains that the average of the current
simplifies into
\begin{eqnarray}
\langle{\cal I}_{j,j+1}\rangle_t && = \sum_{S_2,..,S_{N-1} } \langle S_2,..,S_{N-1} \vert  {\cal I}_{j,j+1} \vert P_t \rangle
\nonumber \\
&& =  2 D_{j,j+1}  \sum_{S_2,..,S_{N-1} } \langle S_2,..,S_{N-1} \vert  \left(\sigma_j^{-}\sigma_{j+1}^{+} 
- \sigma_j^{+}\sigma_{j+1}^{-}  \right) \vert P_t \rangle
\nonumber \\
&& =  2 D_{j,j+1}  \langle  \left( \frac{1+\sigma_{j}^z}{2} \right) \left(\frac{1-\sigma_{j+1}^z}{2}  \right)
-  \left(\frac{1-\sigma_{j}^z}{2} \right)\left(\frac{1+\sigma_{j+1}^z}{2}  \right)  \rangle_t
\nonumber \\
&& =   D_{j,j+1}  \langle  \left( \sigma_{j}^z - \sigma_{j+1}^z \right)  \rangle_t
\label{defaverageti}
\end{eqnarray}
This corresponds to a local Fick law on each bond :
 the averaged current is proportional to the difference of magnetizations with
a prefactor given by the local diffusion coefficient $D_{j,j+1}$. 

\subsection{ Current in the Non-Equilibrium-Steady-State  }

In the non-equilibrium steady state, the magnetizations
\begin{eqnarray}
 \mu_j \equiv \langle \sigma_j^z \rangle_{ness}
\label{dbulk}
\end{eqnarray}
and the currents
\begin{eqnarray}
I_{j,j+1} \equiv \langle{\cal I}_{j,j+1}\rangle_{ness} && =    D_{j,j+1}  (\mu_j-\mu_{j+1})
\label{resi}
\end{eqnarray}
are constrained by the conservation of the current along the chain (Eq \ref{dynm})
\begin{eqnarray}
I= I_{j,j+1} =    D_{j,j+1}  (\mu_j-\mu_{j+1})
\label{iindepj}
\end{eqnarray}

Within the present approach involving the effective dynamics described by the operator $W$ (Eq. \ref{wmaster})
acting on the bulk spins $(S_2,..,S_{N-1})$, 
the magnetizations of the two boundary spins $S_1,S_N$ are fixed to the simple values
\begin{eqnarray}
\mu_1 && = \mu
\nonumber \\
\mu_N && = \mu'
\label{clmu}
\end{eqnarray}
imposed by the boundary-driving (indeed the projector of Eq. \ref{proj0} involves the steady-states of Eq \ref{lindbladspin1n0}
and \ref{lindbladspin2n0} for the boundary spins ).

As a consequence, the current $I$ is simply obtained from the sum of the differences of magnetizations along the chain
\begin{eqnarray}
\mu-\mu' = \sum_{i=1}^{N-1} (\mu_i-\mu_{i+1} )= I   \sum_{i=1}^{N-1} \frac{1}{ D_{j,j+1}  } 
\label{sumdiff}
\end{eqnarray}
leading to the explicit result in each disordered sample
\begin{eqnarray}
I=  \frac{\mu-\mu' }{ \displaystyle \sum_{i=1}^{N-1} \frac{1}{ D_{j,j+1}  } } 
\label{currentres}
\end{eqnarray}
The denominator reads more explicitly in terms of the initial variables (Eq \ref{notaD})
\begin{eqnarray}
  \sum_{k=1}^{N-1}\frac{ 1 }{D_{k,k+1} }
 && = \frac
{ \left(\frac{\Gamma}{2}+2 (\gamma_1+\gamma_2)   \right)^2+4 (h_1-h_2)^2}
{ 4 J_1^2 ( \Gamma+4 (\gamma_1+\gamma_2)) }
+ \frac
{ \left(\frac{\Gamma'}{2}+2 (\gamma_{N-1}+\gamma_N)   \right)^2+4 (h_{N-1}-h_N)^2} 
{ 4 J_{N-1}^2 ( \Gamma'+4 (\gamma_{N-1}+\gamma_N))}
\nonumber \\
&& +\sum_{k=2}^{N-2}
\frac{ (\gamma_k+\gamma_{k+1})^2+(h_k-h_{k+1})^2}
{ 4 J_k^2(\gamma_k+\gamma_{k+1}) }
\label{denoexpli}
\end{eqnarray}

In the limit of large size $N \to +\infty$, this sum will grow extensively in the size $N$ 
\begin{eqnarray}
  \sum_{k=1}^{N-1}\frac{ 1 }{D_{k,k+1} } \oppropto_{N \to +\infty} N \overline{ \left( \frac{ 1 }{D_{k,k+1} }  \right) } + O( N^{\frac{1}{2}})
\label{denoexpliav}
\end{eqnarray}
as long as the disorder-averaged value of the inverse of the local diffusion coefficient converges 
\begin{eqnarray}
   \overline{ \left( \frac{ 1 }{D_{k,k+1} }  \right) } = \overline{ \left(  \frac{ (\gamma_k+\gamma_{k+1})^2+(h_k-h_{k+1})^2}
{ 4 J_k^2(\gamma_k+\gamma_{k+1}) }  \right) } <  +\infty
\label{dinverseav}
\end{eqnarray}
Then the current of Eq. \ref{currentres} will decay as $1/N$ as in the usual Fourier-Fick law.
This is thus completely different from the exponential decay of the current with the system size $N$
 that has been found in the same model in the absence of bulk dephasing \cite{c_lindbladboundary}.
As discussed in the Introduction, this means that the localization properties of the random-field XX chain
survive in the presence of boundary-driving, but do not survive in the presence of bulk-dephasing as expected.

\subsection{ Magnetization profile in the Non-Equilibrium-Steady-State  }

The corresponding magnetization profile reads (Eq \ref{iindepj}) 
\begin{eqnarray}
\mu_j = \frac{ \displaystyle \mu  \left(\sum_{k=j}^{N-1}\frac{ 1 }{D_{k,k+1} } \right) + \mu' \left(  \sum_{k=1}^{j-1}\frac{ 1 }{D_{k,k+1} } \right)}{ \displaystyle \sum_{k=1}^{N-1}\frac{ 1 }{D_{k,k+1} }} 
\label{mjprof}
\end{eqnarray}
that generalizes the usual linear profile of the pure exclusion process without disorder
\begin{eqnarray}
\mu_j^{pure} = \frac{  \mu  (N-j) + \mu' (j-1)}{ (N-1) } 
\label{mjprofpure}
\end{eqnarray}

Eq. \ref{mjprof} means that the magnetization profile in each disordered sample
only displays limited random variations with respect to the usual pure linear profile of Eq. \ref{mjprofpure}.
This is another consequence of the destruction of the localization properties by the bulk dephasing.
Indeed the present nearly-linear magnetization profile 
has to be contrasted with the step magnetization profile that has been found
in the same model in the absence of bulk dephasing \cite{c_lindbladboundary}
and that is expected to occur more generally whenever the coherent bulk dynamics remains localized
\cite{huveneers_mblstep}.

\section{ Two-point correlations }

\label{sec_corre}

After the averaged current and the corresponding magnetization profile studied in the previous section,
it is natural to ask about correlations in the Non-Equilibrium Steady State. 
Indeed, in the field of non-equilibrium classical stochastic processes, long-ranged correlations are expected to be a generic property
of Non-Equilibrium Steady States \cite{derrida}. Remarkably for the pure Symmetric Exclusion Process without disorder, the whole hierarchy of correlation functions has been analyzed \cite{derrida,doucot}. In particular, the two-point correlation follows a very simple form \cite{spohn,speer,derrida}.
For our present effective Exclusion Process with random local diffusion coefficients,
we describe in this section how the two-point correlation can be similarly computed in closed form in each disordered sample.

\subsection{ Dynamics of the two-point correlations  }

For the two-point correlation
\begin{eqnarray}
C_{i,j}(t) \equiv \langle\sigma_i^z\sigma_j^z\rangle_t  
\label{ddefcorre}
\end{eqnarray}
the dynamical equation (Eq \ref{dynA}) reads for $i < j-1$ 
\begin{eqnarray}
\frac{\partial C_{i,j}(t) }{\partial t} && = \langle [\sigma_i^z \sigma_j^z, W] \rangle_t 
\nonumber \\
&&=  \langle ( {\cal I}_{i-1,i} - {\cal I}_{i,i+1} ) \sigma_j^z \rangle_t  +  \langle \sigma_i^z ( {\cal I}_{j-1,j} - {\cal I}_{j,j+1} )  \rangle_t
\nonumber \\
&& = D_{i-1,i} ( C_{i-1,j}(t)  - C_{i,j}(t) )
- D_{i,i+1} (  C_{i,j}(t) - C_{i+1,j}(t)  )
\nonumber \\
&& + D_{j-1,j} ( C_{i,j-1}(t)  - C_{i,j}(t) )
- D_{j,j+1} (  C_{i,j}(t) - C_{i,j+1}(t)    )
\label{dyncorre}
\end{eqnarray}
and for two neighbors $j=i+1$
\begin{eqnarray}
\frac{\partial C_{i,i+1}(t) }{\partial t} && = \langle [\sigma_i^z \sigma_{i+1}^z, W] \rangle_t 
=  \langle  {\cal I}_{i-1,i}   \sigma_{i+1}^z \rangle_t  -  \langle \sigma_i^z  {\cal I}_{i+1,i+2}  \rangle_t
\nonumber \\
&& = D_{i-1,i} ( C_{i-1,i+1}(t)  - C_{i,i+1}(t) )
- D_{i+1,i+2} (C_{i,i+1}(t)  - C_{i,i+2}(t)  )
\label{dyncorren}
\end{eqnarray}

\subsection{ Two-point correlation in the Non-Equilibrium-Steady-State  }

In the Non-Equilibrium-Steady-State, the correlations have thus to satisfy 
linear interpolation formula for fixed $j$
\begin{eqnarray}
C_{i,j} =  \frac{ \displaystyle C_{1,j}  \left(\sum_{k=i}^{j-2}\frac{ 1 }{D_{k,k+1} } \right)
 + C_{j-1,j} \left(  \sum_{k=1}^{i-1}\frac{ 1 }{D_{k,k+1} } \right)}{ \displaystyle \sum_{k=1}^{j-2}\frac{ 1 }{D_{k,k+1} }} 
\ \ \ {\rm for }  \ \ \ 1 \leq i \leq j-1
\label{correreci}
\end{eqnarray}
and for fixed $i$ 
\begin{eqnarray}
C_{i,j} =  \frac{ \displaystyle C_{i,i+1}  \left(\sum_{k=j}^{N-1}\frac{ 1 }{D_{k,k+1} } \right)
 + C_{i,N} \left(  \sum_{k=i+1}^{j-1}\frac{ 1 }{D_{k,k+1} } \right)}{ \displaystyle \sum_{k=i+1}^{N-1}\frac{ 1 }{D_{k,k+1} }} 
\ \ \ {\rm for }  \ \ \ i+1 \leq j \leq N
\label{correrecj}
\end{eqnarray}
while the correlation between two neighbors have to satisfy
\begin{eqnarray}
C_{i,i+1} =  \frac{ \displaystyle C_{1,i+1}  \left(\sum_{k=i+1}^{N-1}\frac{ 1 }{D_{k,k+1} } \right)
 + C_{i,N} \left(  \sum_{k=1}^{i-1}\frac{ 1 }{D_{k,k+1} } \right)}{ \displaystyle \left(\sum_{k=1}^{N-1}\frac{ 1 }{D_{k,k+1} }\right)
-  \frac{ 1 }{D_{i,i+1}} }
\label{correreciipr}
\end{eqnarray}
The correlations with the fixed boundary-spins $\sigma_1^z=\mu$ and $\sigma_N^z=\mu'$  
can be obtained from the magnetization profile of Eq. \ref{mjprof}
\begin{eqnarray}
 C_{1,j}  = \mu \mu_j = \mu \frac{ \displaystyle \mu  \left(\sum_{k=j}^{N-1}\frac{ 1 }{D_{k,k+1} } \right) + \mu' \left(  \sum_{k=1}^{j-1}\frac{ 1 }{D_{k,k+1} } \right)}{ \displaystyle \sum_{k=1}^{N-1}\frac{ 1 }{D_{k,k+1} }} 
\label{correrec1}
\end{eqnarray}
and 
\begin{eqnarray}
 C_{i,N} = \mu' \mu_i = \mu' \frac{ \displaystyle \mu  \left(\sum_{k=i}^{N-1}\frac{ 1 }{D_{k,k+1} } \right) + \mu' \left(  \sum_{k=1}^{i-1}\frac{ 1 }{D_{k,k+1} } \right)}{ \displaystyle \sum_{k=1}^{N-1}\frac{ 1 }{D_{k,k+1} }} 
\label{correrecn}
\end{eqnarray}

Putting everything together, one finally obtains the connected correlation function for $i < j$
\begin{eqnarray}
\label{correijres}
&& C_{i,j}^c  \equiv   C_{i,j} - \mu_i \mu_{j} 
 \\
&& = - (\mu-\mu') \frac{\displaystyle\left(  \sum_{k=1}^{i-1}\frac{ 1 }{D_{k,k+1} } \right)
\left(  \sum_{k=j}^{N-1}\frac{ 1 }{D_{k,k+1} } \right)}{\displaystyle\left(\sum_{k=1}^{N-1}\frac{ 1 }{D_{k,k+1} }\right)^2} 
\left[ \frac{\mu }{ D_{i,i+1}\displaystyle\left(\sum_{k=1}^{N-1}\frac{ 1 }{D_{k,k+1} }
-  \frac{ 1 }{D_{i,i+1}} \right) }  -  \frac{\mu' }{ D_{j-1,j}\displaystyle\left(\sum_{k=1}^{N-1}\frac{ 1 }{D_{k,k+1} }
-  \frac{ 1 }{D_{j-1,j}} \right) } \right]
\nonumber
\end{eqnarray}
This formula generalizes the known expressions for the connected correlations in the pure exclusion process $D_{k,k+1}=1$ \cite{spohn,speer}
\begin{eqnarray}
&& [C_{i,j}^c]^{pure}   = - (\mu-\mu')^2 \frac{ (i-1) (N-j) }{ (N-1)^2 (N-2) } 
\label{correijrespure}
\end{eqnarray}

The important property is that any non-equilibrium steady-state $\mu \ne \mu'$ is characterized by correlations that are weak in amplitude for large size $N$
but long-ranged with respect to the positions $i$ and $j$ 
(see the review \cite{derrida} and references therein).
In particular, the variance of the global magnetization of the sample $M_N=\sum_i \sigma_i^z$
\begin{eqnarray}
< M_N^2>-<M_N>^2 && = \sum_i \sum_j ( < \sigma_i^z \sigma_j^z > - < \sigma_i^z>< \sigma_j^z >)
= \sum_i (1- \mu_i^2) + 2 \sum_{i<j} C_{i,j}^c
\label{varmagnetot}
\end{eqnarray}
gets a non-trivial contribution at leading order $N$ from the double summation of the connected correlation \cite{derrida}.

For two neighbors $j=i+1$, Eq. \ref{correijres} simplifies into
\begin{eqnarray}
C_{i,i+1}^c && \equiv C_{i,i+1} - \mu_i \mu_{i+1} =
- (\mu-\mu')^2 \frac{\displaystyle\left(  \sum_{k=1}^{i-1}\frac{ 1 }{D_{k,k+1} } \right)
\left(  \sum_{k=i+1}^{N-1}\frac{ 1 }{D_{k,k+1} } \right)}
{  D_{i,i+1} \displaystyle\left(\sum_{k=1}^{N-1}\frac{ 1 }{D_{k,k+1} }\right)^2 \displaystyle\left(\sum_{k=1}^{N-1}\frac{ 1 }{D_{k,k+1} }
-  \frac{ 1 }{D_{i,i+1}} \right)} 
\label{correreciip}
\end{eqnarray}
This result will be useful in the next section to compute the fluctuations of the integrated current.

\section{ Current fluctuations  }

\label{sec_fluctuat}

For the Simple Symmetric Exclusion Process without disorder, the current fluctuations have been studied in \cite{doucot}.
In particular, the variance of the integrated current can be obtained from conservation rules \cite{doucot}.
In this section, we describe how this method can be adapted in the presence of disorder.

\subsection{ Integrated current $Q_k$ on a given link }

In order to keep the information on the integrated current $Q_{k}$ on the link $(k,k+1)$ during $[0,t]$,
we need to decompose the ket at time $t$ into a sum over the possible values of $Q_k$
\begin{eqnarray}
 \vert P_t  \rangle =\sum_{Q_k}  \vert P_t (Q_k) \rangle 
\label{decompqk}
\end{eqnarray}
and to write the dynamics of these components
\begin{eqnarray}
\frac{\partial \vert P_t (Q_k) \rangle }{\partial t} = W_{k}^+ \vert P_t (Q_k-2)\rangle +W_{k}^- \vert P_t (Q_k+2)\rangle
+ (W- W_{k}^+- W_{k}^-)  \vert P_t (Q_k) \rangle
\label{masterqk}
\end{eqnarray}
where $W$ is the full operator of Eq. \ref{wmaster}, 
and where the contributions corresponding to the increase or the decrease of the integrated current $Q_k$ are
\begin{eqnarray}
W_k^+ && =  D_{k,k+1} \sigma_k^{-}\sigma_{k+1}^{+} 
\nonumber \\
W_k^- && =  D_{k,k+1}  \sigma_k^{+}\sigma_{k+1}^{-} 
\label{wkpm}
\end{eqnarray}

In particular, the average of the integrated current
\begin{eqnarray}
\langle Q_k\rangle_t =\sum_{Q_k}  Q_k \sum_{S_2,..,S_{N-1} } \langle S_2,..,S_{N-1}  \vert P_t (Q_k)  \rangle  
\label{avqk}
\end{eqnarray}
evolves according to (using the probability conservation $\sum_{S_2,..,S_{N-1} } \langle S_2,..,S_{N-1}  \vert W=0 $)
\begin{eqnarray}
&& \frac{\partial  \langle Q_k\rangle_t  }{\partial t} 
\nonumber \\ && =  \sum_{S_2,..,S_{N-1} } \langle S_2,..,S_{N-1} \vert
\left(   W_{k}^+\sum_{Q_k}  Q_k  ( \vert P_t (Q_k-2)\rangle - \vert P_t (Q_k) \rangle) +W_{k}^-\sum_{Q_k}  Q_k  ( \vert P_t (Q_k+2)\rangle- \vert P_t (Q_k) \rangle )  \right)
\nonumber \\
&& =  \sum_{S_2,..,S_{N-1} } \langle S_2,..,S_{N-1} \vert  ( 2 (W_k^+  -W_k^-) \sum_{Q_k} \vert P_t (Q_k) \rangle
\nonumber \\
&& =   \langle   2 D_{k,k+1} (\sigma_k^{-}\sigma_{k+1}^{+} - \sigma_k^{+}\sigma_{k+1}^{-} ) \rangle_t
= \langle {\cal I}_{k,k+1} \rangle_t
\label{dynqkqk}
\end{eqnarray}
i.e. one obtains the average of the current operator of Eq. \ref{currentop} as it should for consistency.

The average of the square
\begin{eqnarray}
\langle Q_k^2\rangle_t =\sum_{Q_k}  Q_k^2 \sum_{S_2,..,S_{N-1} } \langle S_2,..,S_{N-1}  \vert P_t (Q_k)  \rangle  
\label{avqk2}
\end{eqnarray}
evolves according to
\begin{eqnarray}
&& \frac{\partial  \langle Q_k^2\rangle_t  }{\partial t} 
\nonumber \\
&& =  \sum_{S_2,..,S_{N-1} } \langle S_2,..,S_{N-1} \vert
\left(   W_{k}^+\sum_{Q_k}  Q_k^2  ( \vert P_t (Q_k-2)\rangle - \vert P_t (Q_k) \rangle) +W_{k}^-\sum_{Q_k}  Q_k^2  ( \vert P_t (Q_k+2)\rangle- \vert P_t (Q_k) \rangle )  \right)
\nonumber \\
&& = \sum_{Q_k} \sum_{S_2,..,S_{N-1} } \langle S_2,..,S_{N-1} \vert  \left( 4 (W_k^+ + W_k^-) +4 (W_k^+  -W_k^-) Q_k  \right)\vert P_t (Q_k) \rangle
\nonumber \\
&& = 4 D_{k,k+1}\langle  (\sigma_k^{-}\sigma_{k+1}^{+} + \sigma_k^{+}\sigma_{k+1}^{-} ) \rangle_t 
+  2 \langle  {\cal I}_{k,k+1}   Q_k  \rangle_t
\label{dynqkcarre}
\end{eqnarray}
The first term can be written in terms 
of the two-point magnetization-correlation (using Eq. \ref{defaveraget} and Eq. \ref{defaverageti})
\begin{eqnarray}
\langle (\sigma_k^{-}\sigma_{k+1}^{+} + \sigma_k^{+}\sigma_{k+1}^{-} )\rangle_t 
&& =   \sum_{S_2,..,S_{N-1} } \langle S_2,..,S_{N-1} \vert  \left(\sigma_k^{-}\sigma_{k+1}^{+} 
+ \sigma_k^{+}\sigma_{k+1}^{-}  \right) \vert P_t \rangle
\nonumber \\
&& =  \langle  \left( \frac{1+\sigma_{k}^z}{2} \right) \left(\frac{1-\sigma_{k+1}^z}{2}  \right)
+  \left(\frac{1-\sigma_{k}^z}{2} \right)\left(\frac{1+\sigma_{k+1}^z}{2}  \right)  \rangle_t
\nonumber \\
&& =    \langle  \frac{1-\sigma_{k}^z\sigma_{k+1}^z}{2}   \rangle_t
\label{defaveragetkin}
\end{eqnarray}

The second term of Eq. \ref{dynqkcarre} involves
the correlation between the current ${\cal I}_{k,k+1}  $ and the integrated current $Q_k$.
Since Eq. \ref{dynqkqk} yields
\begin{eqnarray}
\frac{\partial  \langle Q_k\rangle_t ^2 }{\partial t} && = 2 \langle Q_k\rangle_t \frac{\partial  \langle Q_k\rangle_t  }{\partial t} = 
 2 \langle Q_k\rangle_t  \langle {\cal I}_{k,k+1} \rangle_t
\label{dynqkqk2}
\end{eqnarray}
one obtains from the difference with Eq. \ref{dynqkcarre} 
that the dynamics of the fluctuation of the integrated current $Q_k$ involves the connected correlation
the current ${\cal I}_{k,k+1}  $ and the integrated current $Q_k$
\begin{eqnarray}
F_k(t) \equiv \frac{\partial  (\langle Q_k^2\rangle_t  - \langle Q_k\rangle_t ^2 ) }{\partial t} && = 
  2 D_{k,k+1}\langle ( 1-\sigma_{k}^z\sigma_{k+1}^z ) \rangle_t 
+  2 ( \langle  {\cal I}_{k,k+1}   Q_k  \rangle_t  - \langle {\cal I}_{k,k+1} \rangle_t  \langle Q_k\rangle_t  ) 
\label{fktdef}
\end{eqnarray}

Using again Eq. \ref{defaveraget} and Eq. \ref{defaverageti},
one may rewrite Eq. \ref{dynqkcarre} in terms of connected correlations between 
the integrated current $Q_k$ on the link $(k,k+1)$ and the magnetizations $(\sigma_k^z,\sigma_{k+1})$
of the two spins connected to the link
\begin{eqnarray}
F_k(t)  && = 
 2 D_{k,k+1}\langle ( 1-\sigma_{k}^z\sigma_{k+1}^z ) \rangle_t 
\nonumber \\
&& +  2 D_{k,k+1}\left[  ( \langle \sigma_{k}^z  Q_k  \rangle_t  - \langle  \sigma_{k}^z\rangle_t  \langle Q_k\rangle_t)
-  (\langle \sigma_{k+1}^z  Q_k  \rangle_t  - \langle\sigma_{k+1}^z \rangle_t  \langle Q_k\rangle_t  ) \right]
\label{fkt}
\end{eqnarray}

For the special cases of the boundary links $k=1$ and $k=N-1$
involving the fixed spins $\sigma_1^z \to \mu$ and $\sigma_{N}^z \to \mu'$, this simplifies into
\begin{eqnarray}
F_1(t) && = 
  2 D_{1,2} ( 1- \mu \langle\sigma_{2}^z  \rangle_t )
\nonumber \\
&& +  2 D_{1,2}\left[  0 
-  (\langle \sigma_{2}^z  Q_1  \rangle_t  - \langle\sigma_{2}^z \rangle_t  \langle Q_1\rangle_t  ) \right]
\label{fkt1}
\end{eqnarray}
and $k=N-1$
\begin{eqnarray}
F_{N-1}(t) && = 2 D_{N-1,N} ( 1- \langle \sigma_{N-1}^z \rangle_t \mu' ) 
\nonumber \\
&& +  2 D_{N-1,N}\left[  ( \langle \sigma_{N-1}^z  Q_{N-1}  \rangle_t  - \langle  \sigma_{N-1}^z\rangle_t  \langle Q_{N-1}\rangle_t)
-  0 \right]
\label{fktn}
\end{eqnarray}

\subsection{ Comparison of the fluctuations on the different links  }

The integrated currents $Q_{k-1}$ and $Q_k$ on two neighboring links $(k-1,k)$ and $(k,k+1)$ are closely related
since the total change of magnetization of the spin $\sigma_k$ between them
is given by their difference 
\begin{eqnarray}
\sigma_{k}^z(t) - \sigma_{k}^z (t=0) =   Q_{k-1} - Q_k  
\label{magneinte}
\end{eqnarray}
In particular, since this difference remains bounded, the fluctuations $F_k(t)$ introduced above
 will become independent of $k$ and independent of time in the steady-state reached at large time
\begin{eqnarray}
F_k(t) \opsimeq_{t \to +\infty} F
\label{Flimit}
\end{eqnarray}
and the goal is to compute this limit from observables in the steady-state.

From the structure of the system (Eqs \ref{fkt}, \ref{fkt1}, \ref{fktn}),
it is clear that Eq \ref{magneinte} will allow to simplify the following sum
\begin{eqnarray}
\sum_{k=1}^{N-1} \frac{F_k(t) }{ 2 D_{k,k+1}}  && = 
\sum_{k=1}^{N-1} \langle ( 1-\sigma_{k}^z\sigma_{k+1}^z ) \rangle_t 
\nonumber \\
&& +  \sum_{k=2}^{N-1}     ( \langle \sigma_{k}^z  (Q_k- Q_{k-1} ) \rangle_t  - \langle  \sigma_{k}^z\rangle_t  \langle(Q_k- Q_{k-1} )\rangle_t)
\nonumber \\
&&=  1+  \sum_{k=2}^{N-1} \langle\sigma_{k}^z  \rangle_t^2
- \sum_{k=1}^{N-1} \langle \sigma_{k}^z\sigma_{k+1}^z  \rangle_t 
\nonumber \\
&& +  \sum_{k=2}^{N-1}     ( \langle \sigma_{k}^z(t)  \sigma_{k}^z (t=0)\rangle  
- \langle  \sigma_{k}^z(t)\rangle \langle (\sigma_{k}^z (t=0)\rangle )
\label{fktsum}
\end{eqnarray}

\subsection{ Fluctuation $F$ in the Non-Equilibrium-Steady-State  }

In the large-time limit $t \to +\infty$, the time-auto-correlation of the last line of Eq. \ref{fktsum}
can be neglected,
so that
the common value $F$ of the fluctuations (Eq. \ref{Flimit})
can be computed from the knowledge of the magnetization and the two-point correlation
in the steady state
\begin{eqnarray}
 \frac{F}{2} \left(  \sum_{k=1}^{N-1}\frac{ 1  }{  D_{k,k+1}  }\right)    && = 
 1+  \sum_{k=2}^{N-1} \langle\sigma_{k}^z  \rangle_{ness}^2
- \sum_{k=1}^{N-1} \langle \sigma_{k}^z\sigma_{k+1}^z  \rangle_{ness}
\label{ffinal}
\end{eqnarray}

In terms of the magnetizations $\mu_j$ (Eq. \ref{mjprof}) with the boundary conditions $\mu_1=\mu$ and $\mu_N=\mu'$
and of the connected two-point correlation $C^c_{i,i+1}$ (Eq. \ref {correreciip}), the fluctuation $F$ reads
\begin{eqnarray}
 \frac{F}{2} \left(\sum_{k=1}^{N-1} \frac{1 }{  D_{k,k+1}} \right) && = 
1 - \frac{\mu^2+(\mu')^2 }{2}+ \frac{1}{2} \sum_{i=1}^{N-1}  (\mu_i-\mu_{i+1})^2
- \sum_{i=1}^{N-1} C_{i,i+1}^c 
\label{fnesssum}
\end{eqnarray}
Using the difference of magnetizations between consecutive spins (Eq. \ref{iindepj})
 in terms of the current $I$ of Eq. \ref{currentres},
the first sum simplify into
\begin{eqnarray}
  \frac{1}{2} \sum_{i=1}^{N-1}  (\mu_i-\mu_{i+1})^2
=  \frac{I^2}{2} \sum_{i=1}^{N-1} \frac{1}{D^2_{i,i+1}}
=  \frac{(\mu-\mu')^2}{2 \displaystyle\left(\sum_{k=1}^{N-1}\frac{ 1 }{D_{k,k+1} }\right)^2    } \sum_{i=1}^{N-1} \frac{1}{D^2_{i,i+1}}
\label{fness1}
\end{eqnarray}
while the sum of the connected correlation of Eq. \ref{correreciip}
reads
\begin{eqnarray}
- \sum_{i=1}^{N-1} C_{i,i+1}^c 
=\frac{  (\mu-\mu')^2}{\displaystyle\left(\sum_{k=1}^{N-1}\frac{ 1 }{D_{k,k+1} }\right)^2}
 \sum_{i=2}^{N-2}  \frac{\displaystyle\left(  \sum_{k=1}^{i-1}\frac{ 1 }{D_{k,k+1} } \right)
\left(  \sum_{k=i+1}^{N-1}\frac{ 1 }{D_{k,k+1} } \right)}
{  D_{i,i+1}  \displaystyle\left(\sum_{k=1}^{N-1}\frac{ 1 }{D_{k,k+1} }
-  \frac{ 1 }{D_{i,i+1}} \right)} 
\label{fness2}
\end{eqnarray}
so that the final result for the fluctuation $F$ reads in terms of the boundary magnetizations $(\mu,\mu')$
and in terms of the random diffusion coefficients $D_k$
\begin{eqnarray}
F  = 
\frac{2 - \mu^2-(\mu')^2 }{ \displaystyle\left(  \sum_{k=1}^{N-1}\frac{ 1 }{D_{k,k+1} } \right)}
+ \frac{(\mu-\mu')^2}{ \displaystyle\left(\sum_{k=1}^{N-1}\frac{ 1 }{D_{k,k+1} }\right)^3    }
\left[  \sum_{i=1}^{N-1} \frac{1}{D^2_{i,i+1}}
+2
 \sum_{i=2}^{N-2}  \frac{\displaystyle\left(  \sum_{k=1}^{i-1}\frac{ 1 }{D_{k,k+1} } \right)
\left(  \sum_{k=i+1}^{N-1}\frac{ 1 }{D_{k,k+1} } \right)}
{  D_{i,i+1}  \displaystyle\left(\sum_{k=1}^{N-1}\frac{ 1 }{D_{k,k+1} }
-  \frac{ 1 }{D_{i,i+1}} \right)} \right]
\label{fness}
\end{eqnarray}
that generalizes the one obtained for the pure exclusion process $D_{k,k+1}=1$ \cite{doucot}.
For large size $N$, the fluctuation $F$ is of order $1/N$ as the averaged current of Eq. \ref{currentres}
as a consequence of the diffusive nature of the effective dynamics.

\section{  Conclusion}

\label{sec_conclusion}

In this paper, we have studied the Lindblad dynamics of the XX quantum chain with random fields $h_j$ in the presence of two types of dissipative processes, namely dephasing in the bulk and magnetization-driving at the two boundaries. We have focused on the regime of strong disorder in the random fields \cite{prosen_mbl}, or in the regime of strong bulk-dephasing \cite{cai_barthel,znidaric_relaxation}, where the effective dynamics can be mapped via degenerate second-order perturbation theory in the couplings $J_j$
onto a classical Simple Symmetric Exclusion Process with quenched disorder in the diffusion coefficient associated to each bond. We have then studied the properties of the corresponding Non-Equilibrium-Steady-State in each disordered sample between the two reservoirs by extending the methods that have been previously developed for the classical exclusion model without disorder. We have given explicit results for the magnetization profile, for the two-point correlations, for the mean current and for the current fluctuations in terms of the random fields and couplings defining the disordered sample.

As expected, these results are completely different from the transport properties of the same model in the absence of bulk dephasing
\cite{c_lindbladboundary}, where the quantum coherence of the bulk dynamics
 maintains the localized character via a step-magnetization profile and an exponentially decaying current with the system size \cite{c_lindbladboundary}.


\begin{thebibliography}{99}



\bibitem{50years}
``50 years of Anderson Localization'', E. Abrahams Ed, World Scientific (2010).


\bibitem{revue_huse}
R. Nandkishore and D. A. Huse, Ann. Review of Cond. Mat. Phys. 6, 15 (2015).

\bibitem{revue_altman}
 E. Altman and R. Vosk, Ann. Review of Cond. Mat. Phys. 6, 383 (2015).

\bibitem{revue_vasseur}
S. A. Parameswaran, A. C. Potter and R. Vasseur, arxiv:1610.03078.

\bibitem{revue_imbrie}
J. Z. Imbrie, V. Ros and A. Scardicchio, arxiv:1609.08076.

\bibitem{review_mblergo}
D.J. Luitz and Y.B. Lev, arxiv:1610.08993.

\bibitem{review_rare}
K. Agarwal {\it et al}, arxiv:1611.00770.



\bibitem{open}
H.P. Breuer and F. Petruccione, " The theory of open quantum systems", Oxford University Press (2006).

\bibitem{horvat}
M. Znidaric and M. Horvat, Eur. Phys. J. B 86, 67 (2013).


\bibitem{garrahan_mbl}
E. Levi, M. Heyl, I. Lesanovsky and J.P. Garrahan, Phys. Rev. Lett. 116, 237203 (2016) ; \\
B. Everest, I. Lesanovsky, J.P. Garrahan, and E. Levi, arxiv:1605.07019.

\bibitem{prosen_mbl}
M.V. Medvedyeva, T. Prosen and M. Znidaric, Phys. Rev. B 93, 094205 (2016).

\bibitem{altman_mbl}
M.H. Fischer, M. Maksymenko and E. Altman, Phys. Rev. Lett. 116, 160401 (2016)

\bibitem{znidaric_mblsubdiff}
M. Znidaric, A. Scardicchio and V.K. Varma, Phys. Rev. Lett. 117, 040601 (2016).

\bibitem{znidaric_mbldeph}
M. Znidaric, J.J. Mendoza-Arenas, S.R. Clark and J. Goold, arxiv:1609,09367.

\bibitem{huveneers_mblstep}
W. De Roeck, A. Dhar, F. Huveneers and M. Schuetz, arxiv:1606.06076.

\bibitem{c_lindbladboundary}
C. Monthus, arxiv:1701.02102.


\bibitem{casati_step}
G. Benenti, G. Casati, T. Prosen and D. Rossini, Eurphys. Lett. 85, 37001 (2009) ; \\
G. Benenti, G. Casati, T. Prosen, D. Rossini and M. Znidaric, Phys. Rev. B 80, 035110 (2009).

\bibitem{znidaric_deph}
M. Znidaric, New J. Phys. 12, 043001 (2010).

\bibitem{znidaric_trans}
M. Znidaric, Pramana J. Phys. 77, 781 (2011)

\bibitem{znidaric_heisen}
M. Znidaric, J. Stat. Mech. 2011, P12008 (2011)

\bibitem{hubbard}
T. Prosen and M. Znidaric,Phys. Rev. B 86, 125118 (2012).



\bibitem{clark_heat}
J.J. Mendosa-Arenas, S. Al-Assam, S.R. Clark and D. Jaksch, J. Stat. P07007 (2013).




\bibitem{clark_step}
J.J. Mendosa-Arenas, T. Grujic, D. Jaksch and S.R. Clark, Phys. Rev. B 87, 235130 (2013).

\bibitem{prosen_step}
Z. Lenarcic and T. Prosen, Phys. Rev. E 91, 030103(R) (2015).

\bibitem{dragi_trans}
G.T. Landi and D. Karevski, Phys. Rev. B 91, 174422 (2015).



\bibitem{prosen_third}
T. Prosen, New J. Phys. 10, 043026 (2008).

\bibitem{znidaric_solvable1}
M. Znidaric, J. Stat. Mech. 2010, L05002 (2010).

\bibitem{znidaric_solvable2}
M. Znidaric, J. Phys.A: Math. Theor. 43 415004, (2010).

\bibitem{znidaric_solvable3}
M. Znidaric, Phys. Rev.E 83, 011108 (2011).


\bibitem{dragi_ex}
D. Karevski, V. Popkov, and G.M. Schutz, Phys. Rev. Lett. 110, 047201 (2013)

\bibitem{dragi_twist}
V. Popkov, D. Karevski and G.M. Schutz, Phys. Rev. E 88, 062118 (2013)

\bibitem{prosen_mpsolu}
T. Prosen, J. Phys. A: Math. Theor. 48, 373001 (2015).

\bibitem{dragi_mps}
D. Karevski, V. Popkov and G.M. Schutz, arxiv:1612.03601.

\bibitem{derrida}
B. Derrida, JSTAT P07023 (2007).

\bibitem{cai_barthel}
Z. Cai and T. Barthel, Phys. Rev. Lett 111, 150403 (2013).

\bibitem{znidaric_relaxation}
M. Znidaric, Phys. Rev. E 92, 042143 (2015).

\bibitem{kollath}
B. Sciolla, D. Poletti and C. Kollath, Phys. Rev. Lett. 114, 170401 (2015).





\bibitem{garrahan_metastability}
K. Macieszczak, M. Guta,  I. Lesanovsky and J.P. Garrahan, Phys. Rev. Lett. 116, 240404 (2016); \\
D.C. Rose, K. Macieszczak, I. Lesanovsky and J.P. Garrahan, Phys. Rev. E 94, 052132 (2016).

\bibitem{cai}
Z. Cai, C. Hubig and U. Schollwock, arxiv:1609.08518.


\bibitem{garrahan_s}
J.P. Garrahan and I. Lesanovsky, Phys. Rev. Lett. 104, 160601 (2010).

\bibitem{ates_s}
C. Ates, B. Olmos, J.P. Garrahan and I. Lesanovsky, Phys. Rev. A 85, 043620 (2012).

\bibitem{hickey_s}
J.M. Hickey, S. Genway, I. Lesanovsky and J.P. Garrahan, Phys. Rev. A 86, 023609 (2012); \\
J.M. Hickey, S. Genway, I. Lesanovsky and J.P. Garrahan, Phys. Rev. B 87, 184303 (2013).

\bibitem{genway_s}
S. Genway, I. Lesanovsky and J.P. Garrahan, Phys. Rev. E 89, 042129 (2014)


\bibitem{znidaric_slargedev}
M. Znidaric, Phys.Rev.Lett. 112, 040602 (2014).




\bibitem{znidaric_sanomalous}
M. Znidaric, Phys. Rev. B 90, 115156 (2014).

\bibitem{prosen_s}
B. Buca and T. Prosen, Phys. Rev. Lett. 112, 067201 (2014)
T. Prosen and B. Buca, arxiv:1501.06156.

\bibitem{pigeon}
S. Pigeon and A. Xuereb, J. Stat. Mech. (2016) 063203


\bibitem{znidaric_sadditivity}
M. Znidaric, Phys.Rev.E 89, 042140 (2014).



\bibitem{chetrite}
R. Chetrite and K. Mallick, J. Stat. Phys. 148, 480 (2012).






\bibitem{jakob}
M. Jakob and S. Stenholm, Phys. Rev. A 67, 032111 (2003); \\
M. Jakob and S. Stenholm, Phys. Rev. A 69, 042105 (2004).

\bibitem{savona}
E. Mascarenhas, H. Flayac and V. Savona, Phys. Rev. A 92, 022116 (2015).

\bibitem{cirac}
J. Cui, J.I. Cirac and M.C. Banuls, Phys. Rev. Lett. 114, 220601 (2015).


\bibitem{prosen_uniqueness}
T. Prosen, Physica Scripta 86, 058511 (2012).


\bibitem{harris}
R.J. Harris and R.B. Stinchcombe, Phys. Rev. E 70, 016108 (2004).

\bibitem{enaud}
C. Enaud and B. Derrida, Eur. Phys. Lett. 66, 83 (2004).

\bibitem{santen}
R. Juhasz, L. Santen and F. Igloi, Phys. Rev. Lett. 94, 010601 (2005).

\bibitem{igloi}
R. Juhasz, L. Santen and F. Igloi, Phys. Rev. E 74, 061101 (2006).

\bibitem{juhasz}
R. Juhasz, JSTAT P11010 (2011).


\bibitem{spohn}
H. Spohn, J. Phys. A 16, 4275 (1983).

\bibitem{speer}
B. Derrida, J.L. Lebowitz and E.R. Speer, J. Stat. Phys. 107, 599 (2002).

\bibitem{doucot}
B. Derrida, B. Dou\c cot and P.E. Roche, J. Stat. Phys. 115, 717 (2004).

 \end{thebibliography}
\end{document}